\documentstyle[graphicx]{mn}

\title{The X-ray properties of the nearby LINER galaxy, NGC 4736}

\author[T.\,P. Roberts, R.\,S. Warwick \& T. Ohashi]
{T.\,P. Roberts$^{1}$, R.\,S. Warwick$^{1}$ \& T. Ohashi$^{2}$ \\
$^{1}$Department of Physics and Astronomy, University 
of Leicester, University Road, Leicester, LE1 7RH \\
$^{2}$Department of Physics, School of Science, Tokyo Metropolitan University,
1-1 Minami-Ohsawa, Hachioji, Tokyo 192-03}
\date{}

\def\ro{{\it ROSAT~\/}}
\def\asca{{\it ASCA~\/}}

\def\xmm{{\it XMM~\/}}
\def\axaf{{\it AXAF~\/}}

\def\ergcms{{\rm ~erg~cm^{-2}~s^{-1}}}

\def\ergsec{{\rm ~erg~s^{-1}}}

\def\atpcm{{\rm ~atoms~cm^{-2}}}
\def\ctsec{{\rm ~count~s^{-1}}}

\def\H0{{\rm ~km~s^{-1}~Mpc^{-1}}}

\def\ctpix{{\rm ~count~pixel^{-1}}}

\def\xspnorm{{\rm ~photon~cm^{-2}~s^{-1}~keV^{-1}}}

\def\etal{et al.~\/}
\def\eg{{\it e.g.~\/}}

\def\ie{{\it i.e.~\/}}
\def\cf{{\it c.f.~\/}}
\def\la{\mathrel{\hbox{\rlap{\hbox{\lower4pt\hbox{$\sim$}}}{\raise2pt\hbox{$<$}}}}}
\def\ga{\mathrel{\hbox{\rlap{\hbox{\lower4pt\hbox{$\sim$}}}{\raise2pt\hbox{$>$}}}}}

\def\d25{D$_{25}$}
\def\nh{{$N_{\rm H}~$}}
\def\Ha{{H$\alpha~$}}
\def\hi{H {\small I}$~$}

\def\los{line-of-sight\thinspace}
\def\.25{0.25 keV\thinspace}

\def\aa{A\&A}
\def\aas{A\&AS}
\def\mnras{MNRAS}
\def\apj{ApJ}
\def\apjs{ApJS}
\def\apjl{ApJL}
\def\aj{AJ}
\def\pasj{PASJ}
\def\asr{AdSpR}
\def\araa{ARA\&A}

\begin{document}

\maketitle

\begin{abstract}

NGC 4736 is a nearby Sab spiral galaxy, hosting one of the closest
examples of a LINER nucleus.  We have utilized recent observations by
\ro and \asca to characterise the X-ray properties of this galaxy.
Twelve discrete X-ray sources are detected within the region subtended
by its optical disk, the majority of which are likely to be X-ray
binaries associated with the galaxy.  By far the brightest source in
the X-ray band is positionally coincident with the nucleus of the
galaxy and is spatially resolved into a component with a radial extent
of $\sim 3$ kpc plus a point-like core.  The broad band (0.1--10 keV)
spectrum of this nuclear source is composed of a hard continuum with a
spectral slope characteristic of that observed in classical Seyfert
nuclei ({\it i.e.} power-law photon index, $\Gamma \approx 1.7$), with
thermal emission ($kT = 0.1-0.6$ keV) dominant below 2 keV. An Fe
K$_\alpha$ line may also be present at $\sim 6.8$ keV.  There is no
evidence for X-ray temporal variability on timescales of hours to
years.  A plausible model is that the hard continuum originates in a
near-quiescent active galactic nucleus (with $L_X \sim 6 \times
10^{39} \ergsec$, 0.5--10 keV) embedded in the LINER at the centre of
NGC 4736. However, an alternative explanation, namely that the LINER
is the site of a dense population of X-ray binary sources, cannot be
completely excluded.

\end{abstract}


\section{Introduction}

NGC 4736 (Messier 94) is a nearby, face-on spiral galaxy of morphological 
type Sab, located in a sub-group of the Coma-Sculptor cloud of galaxies 
at a distance of 4.3 Mpc (Tully 1988).  Studies over a range of wavebands 
have helped define the distinctive  morphological properties of this galaxy. 
For example, using optical and \hi observations Bosma, van der Hulst 
\& Sullivan (1977) identify five separate annular regions namely 
(i) a bright central region (the bulge), within a radius, $R < 15''$; 
(ii) an inner spiral structure, $R \sim 15''- 50''$, which
is bounded by a bright inner ring; (iii) an outer spiral structure, 
$R \sim 50'' - 200''$, forming an outer disc; (iv) a zone of low 
surface brightness and (v) a faint outer ring at $R \sim 330''$.

Buta (1988) first identified the bright inner ring as a circumnuclear
star-formation region.  In the H$\alpha$ observations of Pogge (1989),
the ring appears oval with a slight extension along the south-east to
north-west axis, and to be clumpy in nature. The ring structure is
also prominent in \hi (Mulder \& van Driel 1993), in molecular gas
(Gerin, Casoli \& Combes 1991) and in the far infrared (Smith \etal
1991).  Further evidence for an active environment
comes from the 6 cm radio continuum observations of Turner \&
Ho (1994), who observe a series of non-thermal emission sources
coincident with the ring which are probably associated with recent
supernovae (RSN) or supernova remnants (SNR). 

The morphology of the bright central region of NGC 4736 has been
studied in detail using surface photometry in the visual, NIR and \Ha
bands by M{\"o}llenhoff, Matthias \& Gerhard (1995), who found
evidence for a weak nuclear bar of size $\approx 30'' \times 9''$
aligned roughly along a north-south axis.  However, most studies of
the nucleus have focussed on spectroscopic data which reveal the
presence of a low-ionisation nuclear emission-line region (LINER,
Heckman, Balick \& Crane 1980). Here the debate centres on whether the
emission line spectrum is powered by a low-luminosity active galactic
nucleus (LLAGN) or is the result of shock excitation and/or
photoionisation from young, hot stars.  Filippenko \& Sargent (1985)
re-classified the source as a {\it Transition\/} LINER since the
low-ionisation nuclear emission lines are interspersed with emission
lines characteristic of star-formation.  This is confirmed by the
study of Larkin \etal (1998), who investigated the near-infrared
spectra of a sample of LINER galaxies, and found NGC 4736 to show the
strongest Fe{\small II}/Pa$\beta$ line ratio of any of their sample,
suggesting that, at least in this waveband, star-formation is the
dominant process.  In fact, the presence of a LLAGN in NGC 4736 is
called into question on the basis of several observations.  Most
particularly, the absorption-corrected \Ha luminosity in the central
(37 pc $\times$ 60 pc) region of the nucleus may be accounted for by
the presence of just six O6 stars (Taniguchi \etal 1996).  Also, the
far-infrared emission, which is peaked at the position of the nucleus,
may be powered solely by the emission of non-OB stars (Smith \etal
1991; Smith \etal 1994), and excludes the presence of an AGN unless
the absorption of the nucleus was underestimated by the authors.  Thus
it has been suggested by Taniguchi \etal (1996) that NGC 4736 is a
``post-starburst'' galactic nucleus, an example of a LINER created in
the absence of AGN activity.

Despite the arguments cited above, a LLAGN may still exist at the
heart of NGC 4736.  For example, Turner \& Ho (1994) have observed a
strong non-thermal radio continuum source at the position of the
nucleus, which they identify as an AGN candidate.  Also, in the recent
survey of nearby galactic nuclei of Ho, Filippenko \& Sargent (1997),
the central 40 pc $\times$ 80 pc region of the nucleus is classified
as a {\it pure\/} LINER region, which again has been taken as evidence
for an AGN (Ho, Filippenko \& Sargent 1998).  More compellingly, {\it
Hubble Space Telescope\/} UV observations of the nucleus of NGC 4736
(Maoz \etal 1995) reveal a bright point source surrounded by diffuse
emission at the exact optical nucleus position.  This source is seen
to have a point-like companion source of equal brightness at a
distance of $2.5''$ ($\sim 50$ pc) directly north of the first
source. Although both sources may be compact star clusters, an
alternative possibility is that NGC 4736 contains not one but {\it
two\/} supermassive black holes that are in the process of merging.
This idea is discussed by Taniguchi \& Wada (1996), who model the
effect of a galaxy merger between a main galaxy and a nucleated
satellite galaxy.  Such a merger may create a supermassive black hole
binary in the nucleus of the main galaxy, which can trigger a
starburst and bow shocks consistent with what is seen in NGC 4736.  A
galaxy merger may also create the ring structure seen in NGC 4736, the
most famous example of such an event being the ``Cartwheel'' galaxy
(Struck \etal 1996).  In this scenario, the weak LINER spectrum and
low \Ha and far-infrared emission in the nucleus may then be explained
by the post-starburst nucleus containing a LLAGN in a relatively
quiescent state.

One way of progressing the debate concerning the origin of the LINER
in NGC 4736 is to consider the X-ray properties of the source. The
most detailed study of NGC 4736 in the high energy band is that
reported by Cui, Feldkhun \& Braun (1997) who have analysed a deep
pointed observation carried out with the \ro X-ray Telescope/Position
Sensitive Proportional Counter (PSPC). They found evidence for an
extended distribution of hot ($kT \approx 0.3$ keV) gas within the
central few kiloparsec region of NGC 4736, in addition to a compact,
relative low luminosity ($L_X \approx 3 \times 10^{39} \ergsec$, 0.1 -
2 keV) nuclear X-ray source. Clearly, a plausible interpretation is
that this compact X-ray source corresponds to a LLAGN, although other
explanations are possible.  Cui \etal (1997) also report the detection
of a further point-like source offset by $\sim 1'$ from the compact
nuclear source, which is perhaps best explained as a bright X-ray
transient in NGC 4736.

Subsequent to the PSPC observation NGC 4736 has been the subject of
further study in the X-ray regime. Specifically it has been observed
on two occasions with the \ro High Resolution Imager (HRI) and also
once by \asca. In this paper we utilize all the available datasets
from \ro and \asca to further investigate both the discrete X-ray
source population in NGC 4736 and the nature of its central X-ray
source. The next section briefly describes the set of observations and
the techniques we have employed to reduce the data. Section 3 then
focusses on the discrete X-ray sources detected in the various \ro
images of NGC 4736, whereas in section 4 we present the results
pertaining to the spatial, spectral and temporal properties of its
nuclear emission. Finally in section 5 we return to the question of
whether the LINER nucleus of NGC 4736 harbours a LLAGN or is powered
entirely by star-formation processes.

\section{Observations and data reduction}

Details of the \ro and \asca observations of NGC 4736 used in the
present paper are given in Table~\ref{4736obs}.  The preliminary
processing of the \ro PSPC data involved excluding time intervals when
the charged-particle master-veto rate was in excess of $170 \ctsec$.
This resulted in the rejection of 11\% of the raw data, leaving $\sim
85$ ks of ``clean'' data.  For the spatial and temporal analysis the
PSPC data were divided into a soft band (corresponding to PI channels
11--41 and energy range 0.1--0.4 keV) and a hard band (PI channels
52--200 and energy range 0.5--2 keV). Preliminary images in both bands
were produced on a $15'' \times 15''$ spatial grid.

\begin{table*}
\caption{Details of the \ro and \asca observations of NGC 4736}
\centering
\begin{tabular}{lccccc}
Satellite/ & Observation & \multicolumn{2}{c}{Pointing direction} & Exposure & Start date \\ 
instrument(s) & identification & RA(2000) & Dec(2000) & (s) & (year.day)\\ & & & & & \\
\ro PSPC	& rp600050n00	& $12^h50^m50.3^s$	& $41^{\circ}06'35''$	& 94819	& 91.156 \\
\ro HRI	& rh600678n00	& $12^h50^m52.7^s$	& $41^{\circ}07'11''$	& 112910	& 94.341 \\
\ro HRI	& rh600769n00	& $12^h50^m52.7^s$	& $41^{\circ}07'11''$	& 27292	& 94.359 \\
\asca SIS $+$ GIS	& 63020000	& $12^h50^m53.0^s$	& $41^{\circ}07'15''$	& 40000	& 95.145 \\
\end{tabular}
\label{4736obs}
\end{table*}

Two separate \ro HRI observations of NGC 4736 have been performed
(Table~\ref{4736obs}); hereafter we refer to the first (and deeper)
observation as HRIa and the second as HRIb.  Preliminary HRI images
were produced using data from PI channels $3-8$ and a
$8'' \times 8''$ pixel grid.

The \asca data were screened using standard procedures (via the FTOOLS
package), leading to a total of $\sim 27$ ks of ``clean'' data from
each of the SIS and GIS instruments.  Since the spatial resolution of
the \asca X-ray telescope is relatively poor, compared to that of the
{\it ROSAT\/} X-ray telescope, the analysis of the \asca observations
concentrated on the spectral and temporal characteristics of the
source.

X-ray spectra of the bright central X-ray source in NGC 4736 (see
section 3) were obtained from the \ro PSPC and from the \asca SIS and
GIS instruments.  The PSPC spectrum was extracted using a circular
aperture of radius of $2'$ centred on the galaxy nucleus. The
resulting pulse-height data were background subtracted (using data
from a source free-region in the \ro PSPC field) and vignetting
corrected and then rebinned into $\sim 30$ spectral channels
encompassing the full 0.1--2.4 keV bandpass of the PSPC. In the case of
\asca, the REV2 standard datasets were used to take advantage of the
most recent \asca calibration files. The source spectra were derived
from a $3'$ radius region centred on NGC 4736 with the background data
taken from the same observation.  The spectral binning was based on
the requirement of a minimum of 20 counts per bin within a spectral
range of 0.6--10 keV and 0.8--10 keV for the SIS and GIS instruments
respectively.

\section{Discrete X-ray sources associated with NGC 4736}

\begin{figure*}
\centering
\includegraphics[width=8cm]{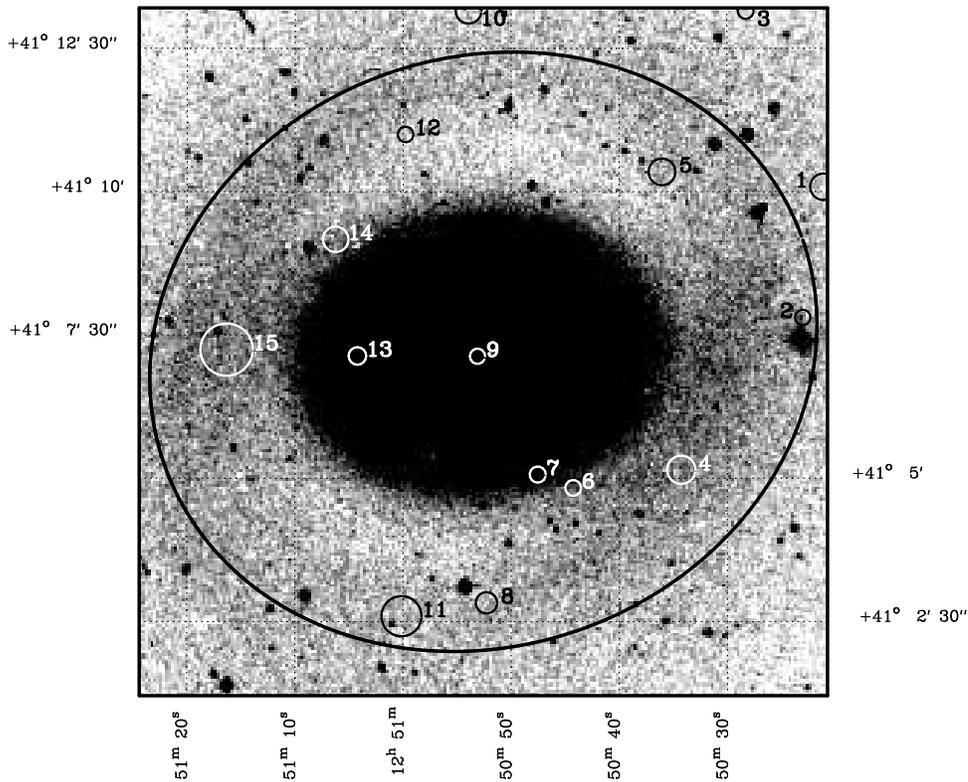}
\caption{The discrete X-ray sources discovered in a $12' \times 12'$ field 
centred on NGC 4736. The position of each source is marked by a circle 
with a radius equal to the positional uncertainty for that source.   
The X-ray data are shown superimposed on the DSS image of NGC 4736 with
the ellipse marking the optical extent of the galaxy (see text).}
\label{4736opt_xsrcs}
\end{figure*}

The central regions of the \ro PSPC and HRI images were searched for
discrete X-ray sources using the PSS software which is part of the
Starlink ASTERIX package (Allan 1995).  Those sources detected in the
{\it ROSAT\/} observations with a significance $ \geq 5\sigma$ by the
PSS algorithm and lying within a $12' \times 12'$ field centred on NGC
4736 are listed in Table~\ref{4736srcs}, with a source designation of
the form ``N4736-X{\it i\/}''. The table also gives for each source
the position and position error (derived preferentially from the HRI
data) and the measured count rate in the soft and hard PSPC bands and
both HRI observations (with a dash signifying a non-detection at the
$5\sigma$ threshold).  A total of 15 discrete sources were detected,
12 of which lie within the elliptical region defined by a major axis
dimension (\d25 diameter) of $11.8'$ and an axial ratio $b/a = 0.87$
(Tully 1988).  Figure~\ref{4736opt_xsrcs} shows the optical image of
the galaxy taken from the Palomar Digitised Sky Survey (DSS) with the
\d25 ellipse and the positions of the X-ray sources overlaid.  The
low-surface brightness gap and the faint outer ring described by
Bosma, van der Hulst \& Sullivan (1977) are clearly visible in the DSS
image; unfortunately the high surface brightness of the three inner
zones renders these regions indistinguishable.  The last column of
Table~\ref{4736srcs} notes possible bright optical counterparts to the
X-ray sources (as discernable in Figure~\ref{4736opt_xsrcs}) and also
whether the source is positionally coincident with either the optical
disc (OD) or the faint outer-ring (OR) of the galaxy.

\begin{table*}
\caption{Discrete X-ray sources discovered within a  $12' \times 12'$ field centred on 
NGC 4736.}
\centering
\begin{tabular}{lcccccccc}
Source	& RA(2000)	& dec(2000)	& Position	& \multicolumn{4}{c}{Count rates ($\times 10^{-3} \ctsec$)}	& Optical \\
 & & & error ($''$)	& SB	& HB	& HRIa	& HRIb	& counterpart? \\
& & & & & & & & \\
N4736-X1	& $12^h50^m20.5^s$	& $41^{\circ}10'03''$	& 13.9	& -	& $0.44 \pm 0.12$	& -	& -	& - \\
N4736-X2	& $12^h50^m23.2^s$	& $41^{\circ}07'48''$	& 8.1	& $1.71 \pm 0.26$	& $1.45 \pm 0.17$	& $1.20 \pm 0.14$	& $1.00 \pm 0.24$	& star? OR  \\
N4736-X3	& $12^h50^m28.5^s$	& $41^{\circ}13'09''$	& 8.3	& $1.38 \pm 0.25$	& $0.76 \pm 0.15$	& $0.58 \pm 0.11$	& -	& - \\
N4736-X4	& $12^h50^m33.7^s$	& $41^{\circ}05'07''$	& 14.8	& -	& $0.45 \pm 0.13$	& -	& -	& OR \\
N4736-X5	& $12^h50^m35.5^s$	& $41^{\circ}10'19''$	& 12.0	& -	& $0.77 \pm 0.15$	& -	& -	& OR \\
N4736-X6	& $12^h50^m44.5^s$	& $41^{\circ}04'49''$	& 8.3	& -	& -	& $0.55 \pm 0.11$	& -	& OD \\
N4736-X7	& $12^h50^m47.8^s$	& $41^{\circ}05'03''$	& 8.4	& -	& $1.66 \pm 0.23$	& $0.60 \pm 0.11$	& -	& OD \\
N4736-X8	& $12^h50^m51.7^s$	& $41^{\circ}02'48''$	& 11.2	& -	& $1.29 \pm 0.18$	& -	& -	& OR \\
N4736-X9	& $12^h50^m53.3^s$	& $41^{\circ}07'07''$	& 8.0	& $76.29 \pm 1.04$	& $86.94 \pm 1.49$	& $49.45 \pm 0.74$	& $54.60 \pm 0.16$	& nucleus \\
N4736-X10	& $12^h50^m53.4^s$	& $41^{\circ}13'08''$	& 13.6	& -	& $0.81 \pm 0.16$	& -	& -	& - \\
N4736-X11	& $12^h50^m59.6^s$	& $41^{\circ}02'34''$	& 21.0	& $1.05 \pm 0.25$	& -	& -	& -	& star? OR \\
N4736-X12	& $12^h51^m00.0^s$	& $41^{\circ}11'00''$	& 8.2	& -	& -	& -	& $1.74 \pm 0.32$	& OR \\
N4736-X13	& $12^h51^m03.7^s$	& $41^{\circ}07'06''$	& 9.2	& -	& $1.05 \pm 0.16$	& -	& -	& OD \\
N4736-X14	& $12^h51^m05.7^s$	& $41^{\circ}09'08''$	& 13.4	& -	& $0.43 \pm 0.13$	& -	& -	& OD \\
N4736-X15	& $12^h51^m15.8^s$	& $41^{\circ}06'38''$	& 27.8	& $0.98 \pm 0.24$	& -	& -	& -	& star(s)? OR \\
\end{tabular}
\label{4736srcs}
\end{table*}

Inspection of Table~\ref{4736srcs} shows that the brightest discrete
X-ray source in NGC 4736 has a count rate at least a factor 30 larger
than any other source and, within the measurement errors, is positionally 
coincident with the optical nucleus of the galaxy.  This ``nuclear''
X-ray source is partially resolved in both the {\it ROSAT\/} PSPC and
HRI observations and is discussed in the next section.  None of the
other sources are bright enough for a detailed X-ray spectral or
temporal analysis to be carried out.  However, it is possible to
comment on the possible nature of some of the sources. For example,
N4736-X2 is the only source, other than the nuclear source, to be
detected in both the PSPC soft and hard bands and also in both HRI
observations. This source lies in the outer ring of NGC 4736 and is
$24''$ north of what appears to be a bright foreground star. However,
for this star to be the source of the X-rays requires either that it
has a large proper motion or that the X-ray positions suffer from an
aspect error (in the form of a rotation). We consider neither of these
a likely scenario. Two further discrete sources spatially coincident
with the outer ring of NGC 4736 (N4736-X11 and N4736-X15) are detected
only in the soft PSPC band.  These may be ``Super-Soft Sources'' (SSS;
see Singh \etal 1995 and references therein), a rare type of object
with a black-body continuum spectrum of temperature of $\sim 30$ eV
and luminosity $^{<}_{\sim} 10^{38} \ergsec$, thought to contain an
accreting white dwarf (van den Heuval \etal 1992).  In fact, if these
sources are at the distance of NGC 4736 and absorbed only by the
foreground Galactic column ($ N_H = 1.4 \times 10^{20} \rm~cm^{-2}$)
then their observed luminosities are of the order $7 \times 10^{36}
\ergsec$ in the 0.1--0.4 keV band, which is not inconsistent with the
SSS hypothesis.  However, the presence of a bright optical object in
the error box of N4736--X11 suggests two further possibilities, namely
that this source is either a foreground object (such as a magnetic CV
or white dwarf) or an ultra-soft background object (such as a
narrow-line Seyfert I; see Brandt, Pounds \& Fink 1995).

Of course, a number of the sources located in the field of NGC 4736
may not be physically associated with the galaxy. For example, the
deep PSPC observation reaches down to a flux threshold of $5 \times
10^{-15} \ergcms$ and using the corrected 0.5--2 keV log $N$ - log $S$
derived by Hasinger \etal (1994), we estimate the source density at
this limit to be $\sim 200$ sources per square degree, or $\sim 5$
sources within the \d25 ellipse of NGC 4736. This is to be compared to
the 11 sources observed (excluding the nuclear source).  However, this
argument does ignore the obscuration due to the \hi associated with
the disc of NGC 4736, an effect which could substantially reduce the
incidence of ``serendipitous'' background sources.  Even if we assume
that the bulk of the sources in the field are associated with NGC
4736, then the implied number of $L_X \sim 10^{37}\ergsec$ sources
associated with this galaxy does not look particularly untoward for a
bright Sab spiral (\cf Read, Ponman \& Strickland 1997).

Using the measured hard band PSPC and HRI count rates we can search
for long term variability provided we assume the sources are all
point-like, (\ie not partially resolved by the HRI) and have a
specific spectral form. Hypothesising that in the 0.5--2 keV energy
range the brightest population of sources in NGC 4736 would be X-ray
binary systems (XRBs), then a reasonable assumption is a thermal
bremsstrahlung spectrum with a mean temperature $kT \approx 1.8$ keV
(Read, Ponman \& Strickland 1997).  We also assume a line of sight
absorption of $3 \times 10^{20} \atpcm$, that is roughly twice
Galactic \nh so as to make some allowance for any absorption intrinsic
to NGC 4736.  On this basis, of the 8 sources investigated, only two
show significant variability between the observations.  Source
N4736-X6 is more than 3 times more luminous in observation HRIa than
in the PSPC observation made 3.5 years earlier (based on an upper
limit to the flux in the latter measurement).  Similarly, N4736-X12
exhibited a factor $\ga 9$ increase between the PSPC and the HRIb
observations (a factor $\ga 3$ change in 18 days is also implied by
the non-detection of this source in the HRIa observation).

\section{X-ray emission from the nuclear source}

\subsection{Spatial properties}

Source N4736-X9 in Table~\ref{4736srcs} is positionally coincident
with the optical nucleus of the galaxy and is, by a substantial
margin, the brightest X-ray source associated with the galaxy.  In the
deep \ro PSPC observation in excess of 10000 counts were accumulated
from this source; similarly 5000 counts were recorded in the HRIa
observation. Thus these datasets allow a fairly detailed spatial
analysis to be performed on, what we refer to hereafter as, the
nuclear X-ray source in NGC 4736.

\begin{figure*}
\centering
\includegraphics[width=8cm]{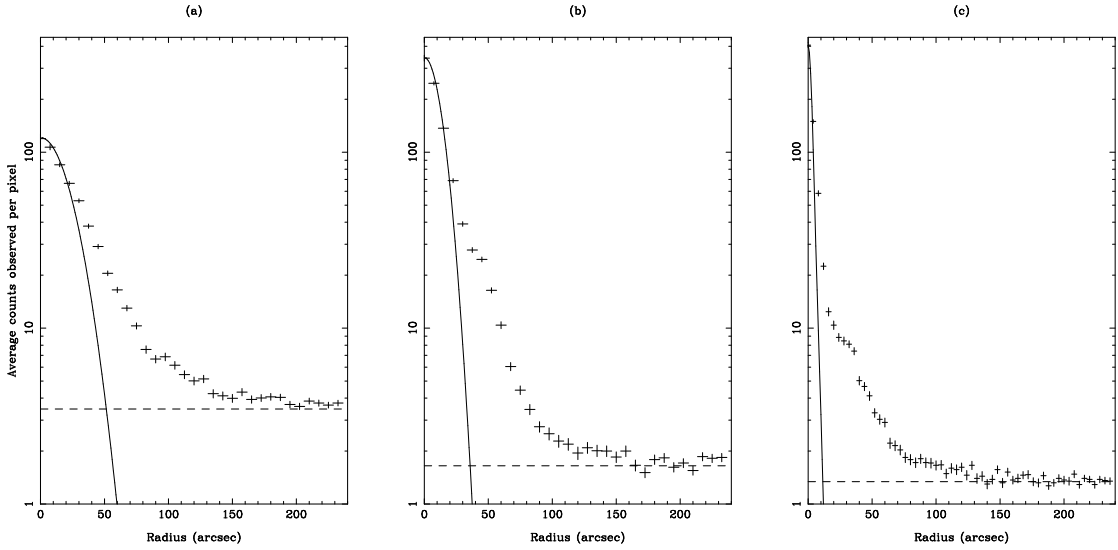}
\caption{Radial profiles of the X-ray emission from the nuclear source in 
NGC 4736 as measured in (a) the PSPC soft band (b) the PSPC hard band
and (c) the HRIa observation.  The PSF for an on-axis source, normalised to 
the peak observed surface brightness, is shown as the solid line.
The nominal detector background is indicated by the horizontal dashed 
line.}
\label{4736radprofs}
\end{figure*}

Inspection of both the \ro PSPC and HRI images suggests that the X-ray
emission from the nuclear source has a spatial extent well in excess
of the predicted point spread function (PSF) for the two
instruments. The magnitude of the effect was investigated by deriving
the radial profiles of the X-ray emission. For this purpose the soft
and hard band PSPC and HRIa data were rebinned onto a $7.5'' \times
7.5''$ and $4'' \times 4''$ pixel grid respectively. X-ray sources
previously detected by PSS (other than the nuclear source itself) were
then masked out, before calculating the radial profiles, centering in
each case on the position of the peak surface brightness.  The
measured radial profiles are shown in Figure~\ref{4736radprofs}, which
also for comparison shows the predicted PSF (the PSF was calculated
from empirical formulae gived by, Hasinger \etal (1995) for the PSPC
and David \etal (1995) for the HRI).  The nominal background was
determined from source-excluded areas outside the region covered by
the profile analysis.

\begin{figure*}
\centering
\includegraphics[width=8cm]{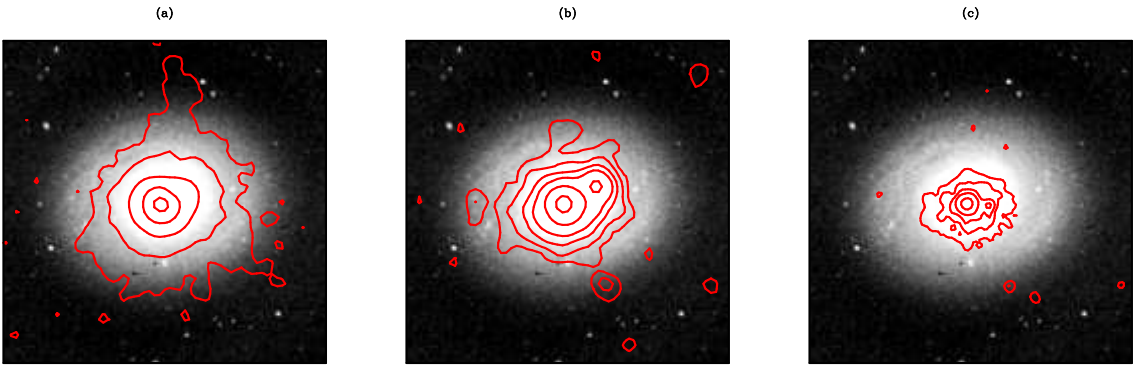}
\caption{Contour maps of the extended X-ray emission region associated
with the nuclear region of NGC 4736.  The three panels correspond to
(a) the  PSPC soft band,  (b) the PSPC hard band and (c) the HRIa
observation.  The contour levels are (a) 5, 10, 30, 65  and $100
\ctpix$;  (b) 3, 5, 10, 20, 60  and $200 \ctpix$; and (c) 3, 6, 10, 20
and $100 \ctpix$.  In each case the field of view is $8' \times 8'$
and the contours are shown  superimposed on the DSS optical image.
The contour maps were smoothed by  convolution with a two dimensional
Gaussian, using $\sigma = 7.5''$ for the  PSPC soft and hard bands and
$\sigma = 4''$ for the HRIa dataset.} 
\label{4736conts}
\end{figure*}

Extended emission, surrounding the central nuclear source, is clearly
seen in both PSPC bands and in the HRI data
(Figure~\ref{4736radprofs}). This emission can be traced out to $\sim
150''$ in all three cases and, in the case the PSPC soft band data,
probably as far as $\sim 200''$ before the signal merges with the
background. At the distance of NGC 4736 the corresponding linear
scales are approximately 3 and 4 kpc respectively.  The total count
rates are 0.138, 0.136 and 0.074 count s$^{-1}$ for the soft and hard
band PSPC and HRIa data respectively; the central point source
contributes 55\%, 64\% and 66\% of the signal in each channel. Since
the signal from the HRI is predominately due to the hard band
(0.5--2.0 keV) emission, the HRI result is consistent with the
resolved fraction being $\sim 10\%$ higher in the soft band.

The morphology of the extended emission in the nuclear region of NGC
4736 is illustrated by the contour plots in Figure~\ref{4736conts},
which are shown overlaid onto the DSS optical image of the galaxy.  In
each case the (fine-grid) images were smoothed by convolution with a
two dimensional Gaussian (we used $\sigma = 7.5''$ for the PSPC soft
and hard bands and $\sigma = 4''$ for the HRIa dataset).
Figure~\ref{4736conts} reveals some dissimilarity between the three
cases.  In the PSPC soft band there is evidence for a low surface
brightness ridge of emission extending to the north of the nucleus
(which, on the basis of the raw image could not be readily attributed
to one or two faint point sources).  This ridge gives rise to the
apparent extension of the soft band radial profile in the
$150''-200''$ range. The central region of the soft band emission is
fairly symmetric whereas in the hard band image there is a distortion
caused by the presence of a relatively bright point source $\sim 1'$
to the north-west of the galactic nucleus. This off-nucleus source
contributes $\sim 7.5\%$ of the total hard band counts\footnote{The
off-nucleus source was not detected by the PSS algorithm. This is
almost certainly due to the use of a larger image pixel size in that
analysis and the position of the source within the confines of the
bright nuclear emission.}.  Potentially the contour map derived from
the HRI data offers the best spatial resolution for studying the
morphology of the extended emission of NGC 4736.  In
Figure~\ref{4736conts}(c) there is no evidence for the off-nucleus
point source detected in the PSPC hard band image; however, a new
point source is seen $32''$ west of the nuclear source.  This source
contributes only $\sim 2\%$ of the total counts and is present at
approximately the same count rate in the HRIb observation performed 18
days later.  Overall the underlying extended emission seen in the HRI
maintains a fairly symmetric form with just a hint of ellipticity in
the contours along an east-west major axis.

Cui, Feldkhun \& Braun (1997) in their earlier detailed analysis of
the \ro PSPC data report the presence of the off-nucleus point source
seen in the hard band image and also note the changed situation in the
HRI (on the basis of a very preliminary look at the HRIb observation).
They postulate that either the two sources are XRBs in the nuclear
region of NGC 4736, or that the source is a foreground object crossing
the \los to the galaxy with a high proper motion.  The displacement
between the X-ray source positions would imply a proper motion of
$\sim 8''$ per year, making the latter scenario highly improbable.
The most likely explanation would seem to be that these are X-ray
transients in the central region of NGC 4736.  Based on the spectral
form discussed earlier for XRBs we estimate the X-ray luminosity of
the PSPC source to be $\sim 3 \times 10^{38} \ergsec$ (in agreement
with Cui \etal 1997) whereas the HRI source has $L_X \sim 9 \times
10^{37} \ergsec$ (0.5--2.0 keV).

\subsection{Spectral properties}

\begin{figure*}
\centering \includegraphics[width=5cm]{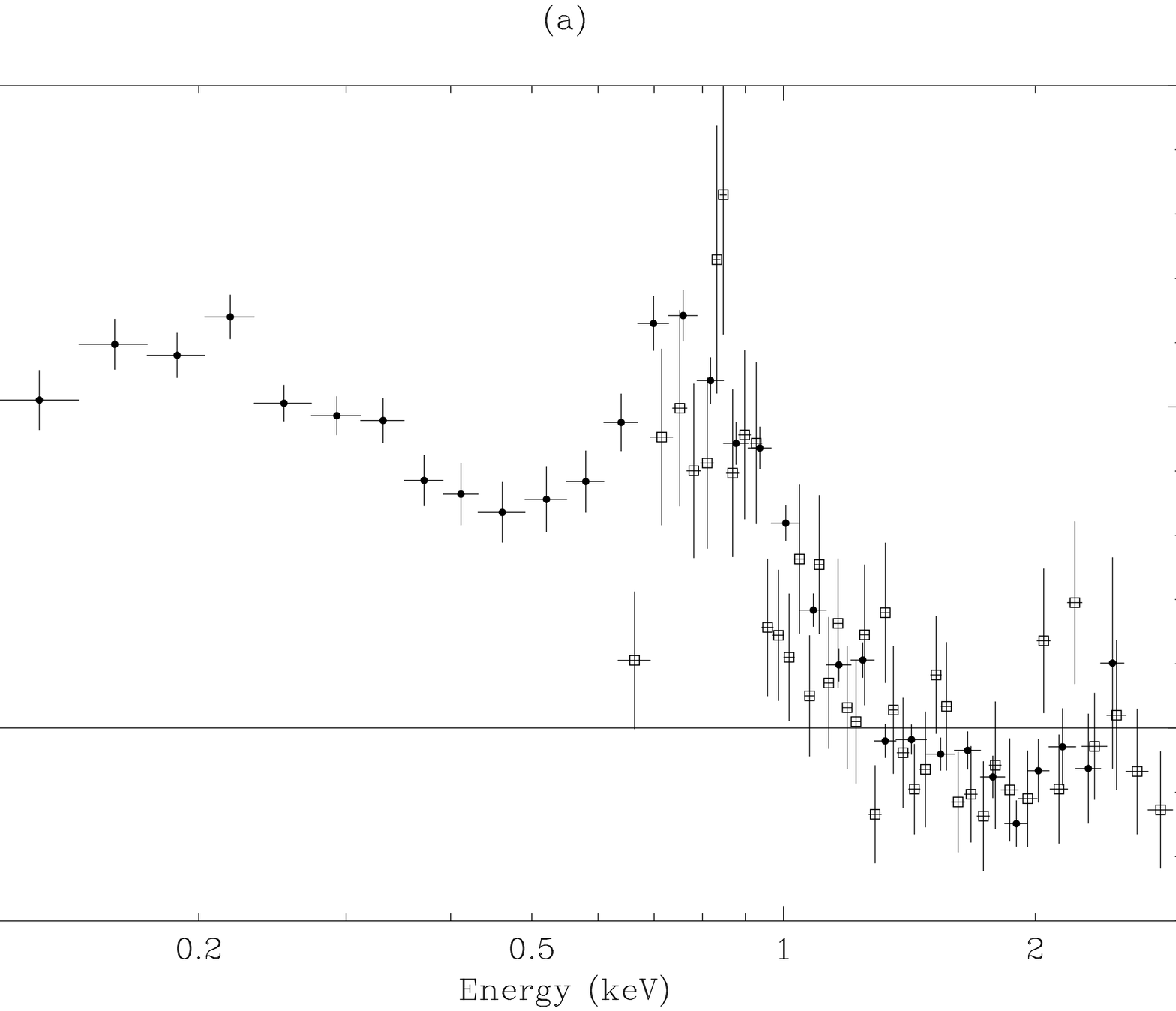}
\hspace*{4cm} \includegraphics[width=5cm]{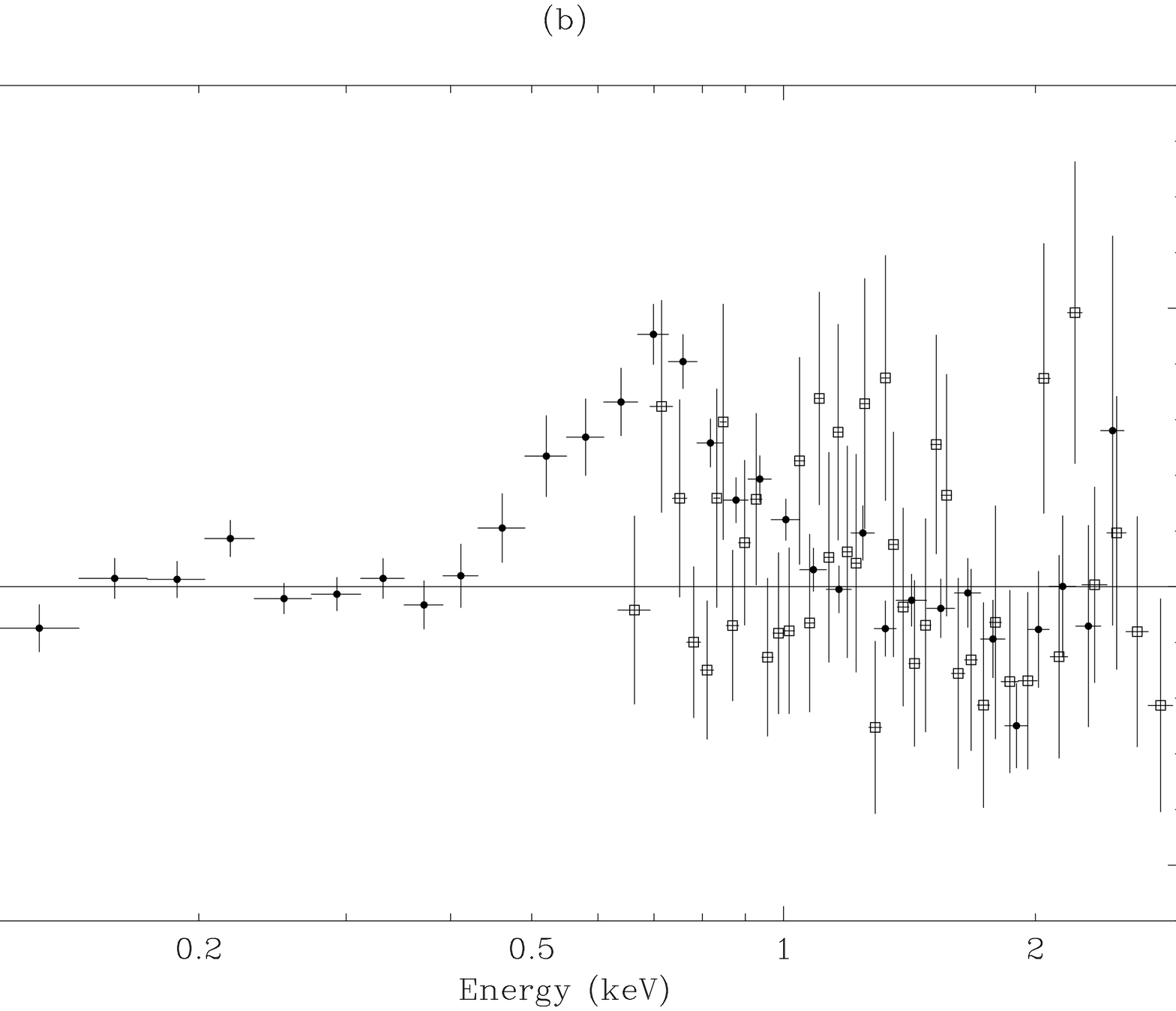}
\includegraphics[width=5cm]{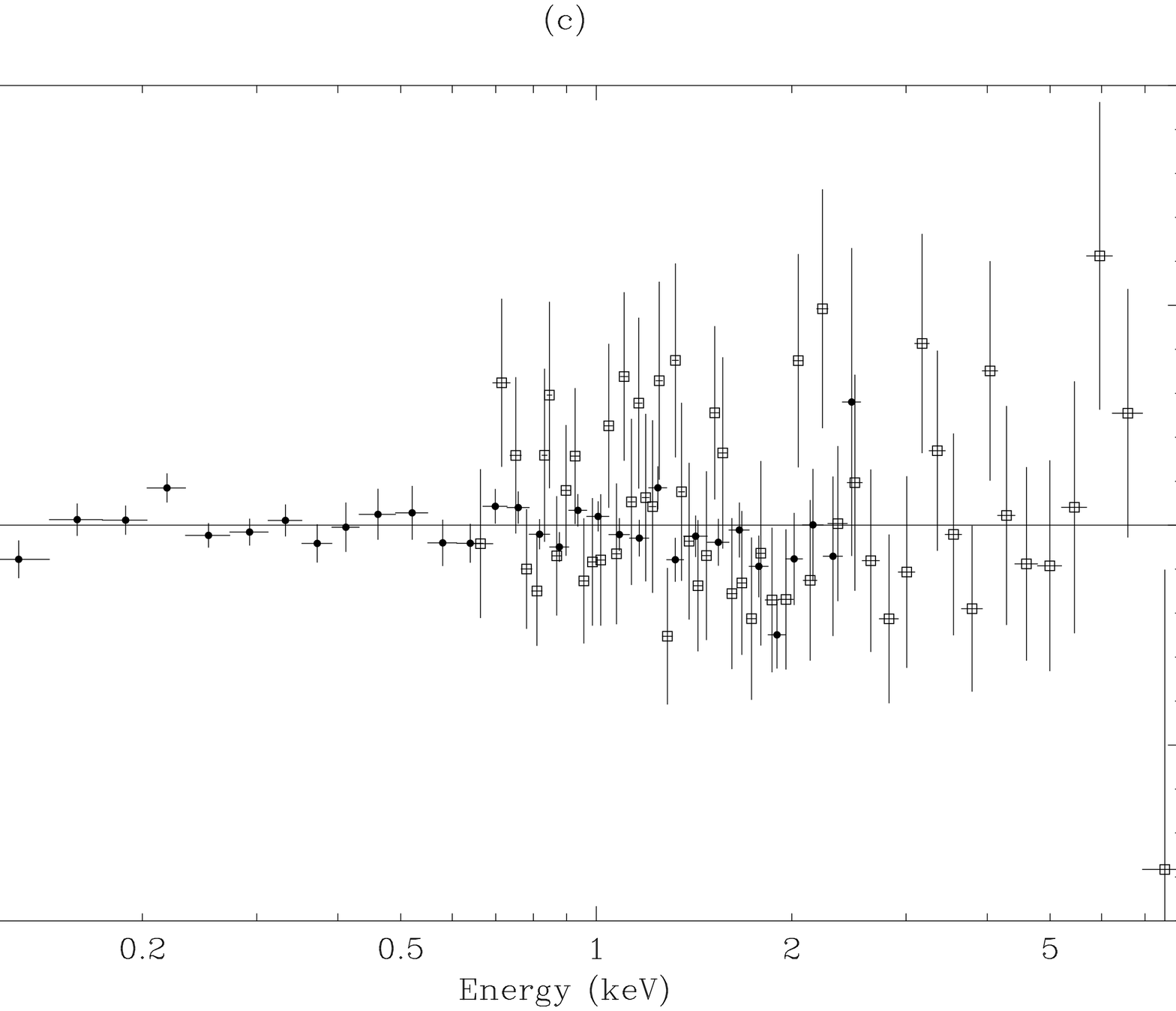}
\caption{(a) The ratio of the \asca (open squares) and \ro PSPC (solid
dots) spectral data to the best-fitting Model-1 prediction,
extrapolated  into the 0.1--3 keV band. (b) The same ratio plot as in
(a) but with the prediction based on Model 3 (except that here the
equivalent width of the $\sim 0.7$ keV Gaussian component is set to
zero). (c) The  same ratio plot as in (b) but with the 0.7 keV
Gaussian component included.  Note that this last plot covers a wider
0.1--8 keV spectral range.  In all three cases only the \asca SIS0
data are shown for clarity.}  \label{4736spec}
\end{figure*}

The \ro PSPC and \asca SIS and GIS instruments provide spectral
coverage in the energy range 0.1--10 keV and together can be used to
examine the broad band spectral characteristics of the nuclear source
in NGC 4736. As a preliminary step, spectral fitting was performed on
the \asca SIS and GIS spectra in the restricted 2--10 keV range
(throughout the spectral analysis the four spectra derived from the
\asca observation were fitted simultaneously using the same
normalisation parameters). A spectral model consisting simply of a
power-law continuum with line of sight absorption due to the Galactic
hydrogen column density ($N_H = 1.38 \times 10^{20} \atpcm$; Stark
\etal 1992), produced a reasonably good fit with a photon index
$\Gamma \sim 1.7$. The replacement of the power-law with a
bremsstrahlung continuum (with $kT \sim 8$ keV) resulted in a similar
match to the data. The inclusion of a narrow ($\sigma = 50$ eV)
Gaussian line component, to represent iron K$_\alpha$ emission, led to
a modest further improvement in the fit ($\Delta\chi^2 \sim 7$). The
best fit parameters (in the case of the power-law continuum) are
listed in Table~\ref{4736allfits} as Model 1, where the errors
correspond to $90\%$ significance for one interesting parameter (see
Yaqoob 1998).  The line energy of $\sim 6.8$ keV is consistent with
the K$_\alpha$ transition of He-like Fe (although the error range
encompasses Fe XX -- XXV, Makishima 1985) and the equivalent width is
high, $\sim 700$ eV, albeit with a large error (since this line is
detected only at $\sim 2.7\sigma$).  When the power-law (or
bremsstrahlung) model is extrapolated below 2 keV, the \asca spectra
reveal an excess of soft X-ray emission (Figure~\ref{4736spec}(a)).
This soft excess flux may be satisfactorily modelled as emission from
a solar abundance Raymond-Smith plasma (see Raymond \& Smith 1977:
Raymond 1992) with a temperature $kT \sim 0.6$ keV (Model 2,
Table~\ref{4736allfits}).

\begin{table}
\caption{The spectral fitting results for the nuclear source.}
\begin{tabular}{lccc}
Parameter & Model 1	 & Model 2	& Model 3 \\ & & & \\ $N_{{\rm
 H}_{fg}}$ $^a$	& 1.38$^f$	  & 1.38$^f$	& 1.38$^f$ \\ & & & \\
 $\Gamma_{PL}$	& 1.75$\pm 0.16$	& 1.70$\pm 0.08$	&
 1.62$^{+0.07}_{-0.08}$ \\ $A_{PL}$ $^b$	&
 4.87$^{+0.98}_{-0.83}$	& 4.44$^{+0.35}_{-0.34}$	&
 3.97$^{+0.26}_{-0.25}$ \\ & & & \\  $E_{K_\alpha}$ $^c$	&
 6.81$^{+0.13}_{-0.28}$	& 6.80$^{+0.13}_{-0.28}$	&
 6.80$^{+0.14}_{-0.32}$ \\ Eq. Width $^d$	&
 679$^{+414}_{-412}$	& 661$^{+397}_{-387}$	& 572$^{+378}_{-357}$
 \\ & & & \\  $kT_{RS1}$ $^c$	& -	&
 0.64$^{+0.05}_{-0.06}$	& 0.65$\pm 0.05$ \\ $A_{RS1}$  $^e$ & -	&
 1.09$^{+0.18}_{-0.19}$	& 1.22$\pm 0.16$ \\ & & & \\  $kT_{RS2}$
 $^c$	& -	& -	& $0.10 \pm 0.01$ \\ $A_{RS2}$  $^e$	&
 -	& -	& 4.33$^{+0.34}_{-0.41}$ \\ & & & \\  E$_{contam}$ $^c
 $ & -	& -	& 0.73$^{+0.02}_{-0.03}$ \\ Eq. Width $^d$	&
 -	& -	& 481$^{+91}_{-89}$ \\ & & & \\  $\chi^2$	&
 73.0	& 175.0	& 223.6	\\ d.o.f.		& 66	& 163	& 191
 \\ & & & \\ 
\end{tabular}

\raggedright

$^a$ $\times 10^{20} \atpcm$ \\ $^b$ $\times 10^{-4} \xspnorm$ \\ $^c$
keV \\ $^d$ eV \\ $^e$ XSPEC units \\ $^f$ parameter value fixed \\

\label{4736allfits}
\end{table}

As a final step in the spectral fitting the PSPC spectrum was also
included in the analysis on the basis of the same spectral model as
employed for the \asca spectra (\ie allowing no spectral or continuum
variability in NGC 4736 between the two observation epochs).  This
demonstrated the need for a further (ultra-soft) spectral component
(see Figure~\ref{4736spec}(a)), which we have modelled in terms of a
second softer Raymond-Smith component.  Again assuming a
solar-abundance plasma we find that a temperature of $kT \sim 0.1$ keV
provides a good match to the spectral data. However, even after the
addition of this new component to the spectral model, the $\chi^2$
residuals in the fit suggested a considerable divergence between the
\asca and PSPC data in the 0.6--1 keV range
(Figure~\ref{4736spec}(b)). After various trials it was found that a
greatly improved $\chi^2$ could be achieved by the addition of a
further narrow Gaussian line component at $\sim 0.7$ keV, fitted only
to the PSPC data (resulting in $\Delta\chi^2 \sim 75$). The resulting
fit parameters are given in Table~\ref{4736allfits} as Model 3 (see
also Figure~\ref{4736spec}(c)).  This line component may represent
some form of contamination of the PSPC spectrum not removed in the
background subtraction process (for example, Cui, Feldkhun \& Braun
(1997) note that solar scattered X-rays are a particular problem in
this PSPC observation) or alternatively may be a by-product (in terms
of the spectral-fitting) of a cross-calibration error between \ro and
\asca.

The above spectral fitting analysis establishes that both a hard
continuum component ( $\Gamma \sim 1.7$ or $kT \sim 8$ keV) and softer
thermal emission, characterised by temperatures in the range $ kT =
0.1-0.6$ keV, are present in the nuclear region of NGC
4736. Unfortunately, apart from the detection of the iron-line
emission, there is insufficient signal to noise and spectral
resolution to define the spectral characteristics in any greater
detail. Thus, although we have assumed solar-abundance in the hot gas,
equally good fits can be obtained if the abundance is dropped to 0.1
solar. Similarly it is not possible to constrain the line of sight
column on the hard continuum (beyond the assumption of Galactic
$N_{\rm H}$) since any intrinsic absorption is effectively masked by
the soft thermal emission. The observations do confirm, however, that
there was no very significant change in the level of the hard
component between the epochs of the \ro PSPC and \asca
observations\footnote{The off-nucleus transient source probably
contributes $\sim 10\%$ to the hard spectral emission detected in the
PSPC observation; thus the flux comparison is reliable to no better
than about $10\%$.}.

It is informative to examine the derived spectral model in relation to
the the spatial structure of the nuclear region established earlier.
Hypothesizing that the extended emission is represented in the
spectral model by the Raymond-Smith components, and the nuclear point
source by the power-law continuum, then we predict, on the basis of
the spectral model, that $\sim 34\%$ of the photons in the 0.1--0.4
keV band and $\sim 60\%$ in the 0.5--2 keV band are associated with
the nuclear point source. By comparison the fractions determined from
the spatial analysis are $55\%$ and $64\%$ (but with the possibility
that the former is somewhat of an over-estimate given the form of the
radial distribution measured in the PSPC soft band, see
Figure~\ref{4736radprofs}(a)). The implication is that the above
division is at least plausible. Conversely, the presence of a bright
point-source in the soft-band PSPC image argues against any
significant intrinsic absorption of the hard continuum, although there
is insufficent information to rule out more ``contrived''
spectral/spatial models.

\subsection{Temporal properties}

An analysis of all of the available observations reveals no evidence
for any significant short-term temporal variability, over periods of
tens of minutes to a few days, in the X-ray flux from the nuclear
source in NGC 4736 (\cf Cui \etal 1997). Also, as noted earlier, the
spectral analysis is consistent with the level of the hard continuum
component remaining unchanged between the \ro PSPC and \asca
observations, which were separated by a time interval of 4 years. A
final check for temporal variability was possible by comparing the
count rates measured for the unresolved central source in the spatial
analysis of the hard-band PSPC data and the HRIa and HRIb
observations. Specifically, assuming that this nuclear point source is
the origin of the hard continuum identified in the spectral analysis,
we have used the PIMMS package (Mukai 1996) to convert the measured
count-rates into fluxes.  The count rates associated with the point
source in the three cases convert to 0.5--2 keV fluxes of $\sim 1.0$,
1.2 and $1.3 \times 10^{-12} \ergcms$ respectively. Any long term
variability implied by these measurements is therefore at an
amplitude $\la 30\%$.

\section{Discussion}

X-ray observations can, in principle, provide crucial information
relating to the question of whether galaxies classified as LINERS
generally harbour LLAGN at their centres.  In the case of NGC 4736,
our analysis of the X-ray emission from its nuclear region confirms
the presence of a bright X-ray point source coincident with the
optical nucleus. We associate this compact source with the hard
continuum evident in the 2--10 keV band.  Modelled as a power-law, the
luminosity of this hard continuum is $\sim 2 \times 10^{39} \ergsec$
in the 0.5--2 keV band, and $\sim 4 \times 10^{39} \ergsec$ in the
2--10 keV range. The measured slope of the power-law corresponds to a
photon spectral index $\Gamma \sim 1.7$, which is fairly typical of
Seyfert 1 galaxies in the 2--10 keV band (e.g. Turner \& Pounds 1989;
Nandra \& Pounds 1994).  Thus, taking into account the lack of any
clear signature of continuum absorption, a possible interpretation of
the NGC 4736 X-ray spectrum is that in the X-ray regime we have a
direct view of a LLAGN which, apparently, is in a near quiescent state
(in that the observed luminosity is a factor $10^{3}-10^{6}$ lower
than is typically of Seyfert nuclei).  The existence of active
galactic nuclei with very low X-ray luminosities is now well
established (e.g. M 81, Ishisaki \etal 1996; NGC 4258, Makashima \etal
1994; NGC 3147, Ptak \etal 1996; M51 Terashima \etal 1998a) and in the
most extreme case reported to date, namely that of NGC 4395, the dwarf
Seyfert nucleus was found to have an $L_X$ of only $\sim 3 \times
10^{38} \ergsec$ (Lira \etal 1998).

In order to confirm the supposition that the LINER nucleus in NGC 4736
contains a LLAGN it is necessary to consider what further clues are
provided by the X-ray data. Support for the presence of a LLAGN could
in principle come from the observation of significant continuum
variability. If LLAGN are the direct low-luminosity analogue of the
classical Seyfert phenomenon then the trend of increasing variability
with decreasing luminosity displayed in Seyfert galaxies (Nandra \etal
1997a) would predict LLAGN to be highly variable X-ray sources.  This
property is not apparent in the sample of LLAGN and LINERs studied by
Ptak \etal (1998), although M 81 and NGC 4395 do vary on a timescale
of $\sim$ weeks.  Ptak \etal (1998) argue that since LLAGN must be accreting
at grossly sub-Eddington rates, they are likely to be powered by an
advection-dominated accretion flow (Narayan \& Yi 1995), a scenario in
which rapid X-ray variability can be suppressed. In any event, it is
not possible to use the absence of X-ray temporal variability in NGC
4736 on timescales of $\sim$ hours to days as a strong argument either
for or against the LLAGN hypothesis.

Potentially the most important clue we have as to the origin of the
hard X-ray emission is provided by the detection of the Fe K$_\alpha$
emission line, with recent results from \asca amply demonstrating the
diagnostic possibilities of such features (\eg Nandra \etal 1997a).
In the case of NGC 4736 we detect (albeit at marginal significance) a
Fe-K line with a high equivalent width ($\sim 600$ eV) at an energy
inconsistent with the fluorescence in cold gas.  The most obvious
explanation of the large equivalent width, by analogy to the situation
pertaining in classical Seyfert 2 galaxies such as NGC 1068 (see
Antonucci 1993), is that the direct \los to the central engine is
completely blocked, perhaps by a molecular torus, but nevertheless we
see a fraction of the nuclear flux as a result of electron scattering.
This scattering occurs in a photoionised medium which extends
sufficiently far from the nucleus, probably along an axial direction,
so as to be visible above the plane of the obscuring medium. Typically
up to a few percent of the nuclear continuum can be redirected into
our line of sight and a substantial Fe-K line feature is superimposed
due to the fluorescence of the scattering medium. Iron line equivalent
widths greater than 1 keV can be produced by this means (\eg Makashima
1985), with the line energy providing a measure of the ionisation
state of the electron scattering region. For example, at least two
{\it high energy} Fe-K components are detected with high equivalent
width in the X-ray spectrum of NGC 1068, indicative of the presence of
a warm scattering medium (Marshall \etal 1993; Ueno \etal 1994;
Iwasawa, Fabian \& Matt 1997).
 
In the case of NGC 4736, the combination of the observed Fe-K feature,
the absence of temporal variability and the lack of any clear
signature of continuum absorption constitute circumstantial evidence
in favour of a scattering scenario (with the implication that the
direct nuclear emission is blocked by a column density well in excess
of $10^{24} \rm~cm^{-2}$). The luminosity of the ``hidden'' LLAGN
might then be close to $10^{41} \ergsec$ (if we assume a scattered
fraction of roughly $3\%$ of the total nuclear continuum). A test of
whether NGC 4736 might harbour such an ``unseen'' type-2 nucleus is
provided by reference to its far-infrared flux. David \etal (1992)
derive a linear relationship between the FIR luminosity (based on 60
and 100 micron fluxes) and the (0.5--4.5 keV) X-ray luminosities of a
sample of normal, starburst and LINER galaxies. Applied to NGC 4736, the
David \etal formula predicts an expected X-ray luminosity of $\sim 5
\times 10^{39} \ergsec$ in close agreement to that actually
observed. In principle this argues against {\it any\/} LLAGN presence
in NGC 4736, but given the scatter in the David \etal correlation, it
remains a reasonable conjecture that a LLAGN at a ``quiescent'' level
of a few times $10^{39} \ergsec$ has escaped detection in other
wavebands. However, the requirements for a much more luminous hidden
LLAGN look implausible. Similarly the observed L$_{\rm [OIII]}/L_X$
ratio for NGC 4736 is much closer to the norm for unobscured Seyfert
nuclei than for galaxies with hidden Seyfert nuclei such as NGC 1068
(Mulchaey \etal 1994). Unfortunately in the quiescent LLAGN scenario
we are left with the problem of how to explain both the observed
energy and equivalent width of the Fe K$_\alpha$ line.  Terashima
\etal 1998b discuss the relevant arguments in the context of the LINER
galaxy NGC 4579, which also has a strong He-like iron Fe-K line in its
\asca spectrum.

There is of course an alternative to the LLAGN hypothesis, namely that
all of X-ray flux emanating from nuclear point source in NGC 4736 is
generated by stellar mass objects in the nucleus of the galaxy.  Given
the relatively hard spectrum, the most likely population would seem to
be low and/or high mass X-ray binaries (XRB) (see Fabbiano 1989;
Makishima \etal 1989).  For example, $\la 40$ XRB concentrated in the
nucleus of the galaxy, powered at near to their Eddington limit by
accretion onto a neutron star or stellar mass black-hole, could
produce both the observed X-ray luminosity and the hard spectral form.
Also in this model, since the observed X-ray flux is due to the
integrated emission of a reasonable number of separate sources, a
substantial amplitude of variability is not expected.  The one major
difficulty with this description is in explaining the apparently large
equivalent width of the Fe K line in NGC 4736, since individual XRB
tend to have Fe K lines with equivalent widths of only a few tens of
eV (Hirano \etal 1987), which is well outside the formal $90\%$
confidence range of the measurement.  As a check of the XRB model we
can utilize the fact that, in the absence of an AGN, the optical
luminosity of a spiral galaxy bulge is often well correlated with its
X-ray luminosity due to its population of XRBs (Canizares, Fabbiano \&
Trinchieri 1987). In the case of NGC 4736 this method predicts, on the
basis of the optical magnitude of the bulge of NGC 4736 ($M_B =
-18.19$; Ho, Filippenko \& Sargent 1997) and Figure 5 of Iyomoto \etal
(1998), an X-ray luminosity of $\sim 1 \times 10^{39} \ergsec$, which
is a factor four below that observed. However, an over-density of
XRBs, particularly in a region characterised by recent intense
starburst activity, perhaps with an XRB population biassed toward high
mass systems, cannot be excluded,

An interesting feature of bright nuclear X-ray source in NGC 4736 is
that over a third of the X-ray emission below 2 keV is resolved with a
radial extent of $\sim 150''$, {\it i.e.} $\sim 3$ kpc.  We associate
this extended emission with the two thermal components identified in
the spectral analysis with temperatures of $0.1$ and $0.65$ keV, which
have a combined luminosity of $\sim 2 \times 10^{39} \ergsec$ in the
0.1--2 keV PSPC band. The observed thermal emission is largely
confined within the bright circumnuclear ring of star-forming regions
identified in \Ha and other wavebands.  A possible origin of this hot
gas is in a starburst event (or events) which have occurred in the
nucleus of NGC 4736.  Hydrodynamic modelling of starbursts (\eg
Suchkov \etal 1994; Tenerio-Tagle \& Mu{\~n}oz-Tu{\~n}{\'o}n 1997) has
been broadly successful in duplicating the results of X-ray
observations that show kpc scale ``plumes'' of hot X-ray emitting gas
escaping from the disc into the halo of galaxies experiencing nuclear
starbursts (\eg NGC 253, Pietsch \& Tr{\"u}mper 1993).  For the hotter
plasma observed we derive a volume emissivity of $\sim 2 \times
10^{-27} \rm~erg~cm^{-3}~s^{-1}$ (assuming spherical symmetry and a
concentration of the emission within a radius of $100''$), an electron
density of $\sim 10^{-2}\rm~cm^{-3}$ and a cooling time of $\sim
10^{9} \rm~ yr$ (Tucker 1975). Hence the hot extended plasma component
may indeed have had its origin in a recent starburst epoch.

Unfortunately, the modelling of the soft X-ray excess observed in NGC
4736 in terms of two specific temperature components may not represent
an accurate picture of the state of the hot plasma in this galaxy.  A
variety of effects such as a departure from non solar-abundances, a
smooth spread of temperature values and non-equilibrium conditions in
the plasma may have combined to mimic a two-temperature spectrum
(given the limited spectral resolution and sensitivity afford by the
combined \ro and \asca data).  Equally well it is possible to
hypothesize that the two Raymond-Smith components may represent two
distinct spatial distributions, for example the hotter plasma could be
located within the disc of the galaxy with the cooler component
forming a more diffuse halo. Unfortunately, the only real evidence for
such a spatial segregation is given by the low-surface brightness
North-South ridge of emission, evident in the soft band but not the
hard band PSPC image.

\section{Conclusion}

There has been considerable recent interest in the identification of
candidate LLAGN in the nuclei of nearby galaxies via X-ray
measurements. From \asca observations we know that many of the sources
established as bona-fide LLAGN have broadly similar spectral
characteristics to those observed in the nuclear source in NGC 4736,
namely a hard continuum, a detectable Fe-K line and a softer, probably
thermal component, which is prominent below 2 keV (see Serlemitsos,
Ptak \& Yaqoob 1996). Unfortunately, despite this apparent conformity,
we are still unable to answer the key question of whether the LINER
nucleus in NGC 4736 contains an X-ray bright LLAGN with complete
certainty. Fortunately, we can be fairly sure that when high
sensitivity X-ray observations with good spectral and spatial
resolution become available from \axaf and \xmm, a definitive answer
will be forthcoming. As in many areas of X-ray astrophysics, the Fe-K
line may well provide the crucial diagnostic.

\section{Acknowledgements}

TPR acknowledges support from PPARC initially in the form of a
research studentship and more recently as a postdoctoral research
associate.  The X-ray spectral data used in this work were obtained
from the Leicester Data Archive Centre (LEDAS) and the HEASARC
facility at Goddard Space Flight Centre, USA. The Digitized Sky Survey
was produced at the Space Telescope Science Institute, under US
Government grant NAG W-2166 from the original National Geographic -
Palomar Sky Survey Plates.


\end{document}